\documentclass[12pt]{article}

\title{Quark matter conductivity  in strong magnetic background\\}
\author{B.Kerbikov\thanks{E-mail:borisk@itep.ru} \\
 State Research
Center\\Institute for Theoretical and Experimental Physics, \\
Moscow, Russia\\{\it Talk at the International Conference CPOD 2010}}
 \date{}

\newcommand{\beq}{\begin{eqnarray}}
 \newcommand{\eeq}{\end{eqnarray}}
  \newcommand{\be}{\begin{equation}}
\newcommand{\ee}{\end{equation}}
 \def\ga{\mathrel{\mathpalette\fun >}}
\def\fun#1#2{\lower3.6pt\vbox{\baselineskip0pt\lineskip.9pt
\ialign{$\mathsurround=0pt#1\hfil ##\hfil$\crcr#2\crcr\sim\crcr}}}

\newcommand{\vek}{\mbox{\boldmath${\rm k}$}}

\newcommand{\veq}{\mbox{\boldmath${\rm q}$}}

\begin{document}

\maketitle
\begin{abstract}
Applying the ideas and methods of condensed matter physics we calculate the
quantum conductivity of quark matter in magnetic field. In strong field quantum
conductivity is proportional to the square root of the field.

\end{abstract}

After a decade of RHIC performance it has become clear that we are encountering
an unusual form of matter. To decipher its properties methods traditional to
condensed matter physics and hydrodynamics turn out to be effective. A very
intriguing effect observed by STAR collaboration at RHIC is the electric
current induced in the direction of the magnetic field. This is now famous
Chiral Magnetic
 Effect -- see \cite{1,2} and references therein. The magnetic field created
 by heavy ion currents  at the collision moment  is huge, $|eB|\ga m^2_\pi \sim 10^{19} G $ \cite{2}. Magnetic field
of the same order or even higher is expected at LHC. It is therefore clear that
the problem of quark matter conductivity in strong magnetic field is an
important albeit a complicated one. In what follows we shall propose a solution
which may be traced back to the well known results of condensed matter physics
\cite{3,4}. Two remarks are in order before we proceed to the issue. First,
 is that in the  present short contribution details of the calculations are omitted. Second, is that here we shall not
try to establish links to the Chiral Magnetic effect.

     Our basic assumptions are the following. We postulate the formation of the
     Fermi surface in dense quark matter and assume that transport coefficients
     including conductivity are  defined by the physical processes occurring in
     the vicinity of this surface. Next we assume that a certain level of
     disorder is present giving rise to the scattering time $\tau $ (in
     condensed matter physics this is the electron elastic scattering time on
     impurities). Following \cite{5} we suppose the quark matter to be in a
     regime of weak  localization \cite{3,4}, i.e., $ \tau_\varphi\gg \tau$, $
     k_F l \sim 1$, where $\tau_\varphi$ is the phase-breaking time, $k_F$ is
     the Fermi momentum, $l$ is the mean free path (at $k_Fl<1$ transition to
     Anderson localization takes place). Therefore we shall describe
     conductivity as  a quantum  process in the background of strong
     fluctuations \cite{5,6}. Is so, two physical parameters play the key role
     -- phase-breaking time $\tau_\varphi$ and the diffusion coefficient $D$
     \cite{3,4,5}. There is an important difference from the typical picture is
     condensed matter physics. Namely, quantum contribution to the conductivity
     is important even  in presence of strong magnetic field. This is a subtle
     point  which will be elucidated in detail in the forthcoming publication.
     We conceive this conclusion to be true in the three-dimensional case
     considered here. Ultra-relativistic ions resemble two-dimensional discs
     and in this case quantum conductivity is infrared divergent \cite{3,4}.

     The expression for the conductivity reads \cite{3,4}
     \be \sigma_{\alpha\beta} =-\lim_{\omega\to 0} \frac{Q_{\alpha\beta}
     (\omega)}{i\omega},\label{1}\ee
     where $Q_{\alpha\beta}(\omega)$ is the electromagnetic response operator
     defined for Matsubara frequencies $\omega_\nu =2\pi \nu T, ~~
     i\omega_\nu\to \omega$. Since we consider the regime of weak localization
     fan diagrams enter into the response operator on equal footing with ladder
     diagrams \cite{3,4}. This means that the response operator has the form

     \be Q_{\alpha\beta} (\omega_\nu) = T \sum_{\varepsilon_n} \int
     \frac{d\veq}{(2 \pi )^3}\frac{d\vek}{(2 \pi )^3} j_\alpha G_1G_2C (\veq,
     \omega_\nu) G_3G_4 j_\beta,\label{2}\ee
     where $G_i (i=1,2,3,4)$ are relativistic Matsubara propagators at finite
     $T$ and $\mu$ \cite{7} with the following arguments:  $ G_1=G (\vek, i
     \tilde \varepsilon_n)$, $G_2 = G (\vek, i\tilde \varepsilon_{n+\nu})$,

 $G_3 = G (\veq-\vek, i\tilde \varepsilon_{-n-\nu})$, $G_4 = G (\veq-\vek, i\tilde \varepsilon_{-n})$,
$\tilde \varepsilon = \varepsilon_n +(2\tau)^{-1} sgn \varepsilon_n, ~
\varepsilon_n =\pi (2n+1) T$. The quantity $C(\veq, \omega_\nu)$ is Cooperon
\cite{3,4} \be C^{-1} (\veq, \omega_\nu) = 4 \pi \nu \tau^2 (\omega_\nu+
D\veq^2 + \tau^{-1}_\varepsilon),\label{3}\ee where $ \nu=\mu k_f/2\pi^2$ is
the relativistic density of states at the Fermi surface, $ D( v^2_F, T,\tau)$
is the diffusion coefficient \cite{3,8}, the factor $4\pi$ is replaced by
$2\pi$ in nonrelativistic case. Due to Cooperon diffusion like pole appears in
the polarization operator. When we impose magnetic field $B$ along  $z$-axis
$D\veq^2$ is replaced by $(Dq^2_z+\Omega(k+1/2))$, where $\Omega=4 e DB$, $e$
is the  absolute value of the quark electric charge, $2D$ replaces the inverse
mass in the cyclotron frequency, $k$ enumerates Landau levels. Next one inserts
into (\ref{2}) a complete set of Landau states and the normalization factor
counting the number of states per unit area of a full Landau level. Calculation
of (\ref{2}) is a  somewhat cumbersome exercise which will  be presented
elsewhere (for nonrelativistic case see \cite{3,4}). Here we list our
approximations and present the results. Only quark contributions are kept in
the propagators $G_i$ (no antiquarks).   Integration over $\vek$ is replaced by
integration around the Fermi surface over $\xi =\sqrt{\vek^2+m^2}-\mu$.
Propagators $G_i$ are taken in the $\tau$ -- approximation (dirty limit
\cite{3}). Current operators $j_\alpha=e\gamma_\alpha$ are expressed via the
corresponding  momenta using Gordon relation. In the  static limit the result
for quantum contribution to the conductivity reads \be \sigma =- 4 e^2 N_f N_c
|eB|D\int \frac{d q_z}{2\pi} \sum_k \frac{1}{Dq^2_z +\Omega (k+1/2)
+\frac{1}{\tau_\varphi}}\label{4}\ee where we have used the equation for the
diffusion coefficient $D= k^2_F \tau/ 3\mu^2$, $\mu=(k^2_F +m^2)^{1/2}$. This
expression is valid in the dirty limit\cite{8}. It comes as not a surprise that
we have retrieved the non-relativistic result \cite{3,4} with only minor
changes (we remind that antiquark contribution is omitted). the negative sign
in (\ref{4}) means that quantum effects result in  negative magnetoresistence
and may drastically suppress the total conductivity.

 Next we have to  estimate the parameters entering into (\ref{4}) in order to
 see what is the value of the magnetic field that kills weak localization. The
 critical field is
 \be |eB_c|\simeq \frac{\pi}{D\tau_\varphi} \gg m^2_\pi,\label{5}\ee
 where for the estimate we used the above expression for the diffusion
 coefficient taken in the chiral limit with $\tau\simeq 1 $ fm and took for the
 phase-breaking time the value $\tau_\varepsilon \simeq 4 $ fm. On the other
 hand the value of the magnetic field at RHIC $|e B| \sim m^2_\pi$ is strong
 enough to guarantee the smallness of the dimensionless parameter $\delta = (4
 |eB| D\tau_\varphi)^{-1}$. In this limit expression (\ref{4}) yields in
 three-dimensional case the square root dependence on the magnetic field,
 $\sigma \sim (|eB|)^{1/2}$ \cite{3,4}.

The author is indebted to A.Varlamov, M.Andreichikov, V.Novikov, M.Vysotsky and
L.Levitov for discussions and comments, Support from RFBR grants
08-02-92496-NTSNIL-a and 10-02093111-NTSNIL-a are gratefully acknowledged.

\end{document}